\documentclass[10pt]{blois07}
\usepackage{graphicx,epsfig}
\usepackage{cite,./mcite}

\begin{document}
\title{High energy QCD beyond the mean field approximation}
\author{Arif~I.~Shoshi\footnote{The author acknowledges financial support
  by the Deutsche Forschungsgemeinschaft under contract 92/2-1.}}
\institute{Fakult\"at f\"ur Physik, Universit\"at Bielefeld, D-33501
  Bielefeld, Germany}
\maketitle

\begin{abstract}
  It has been recently understood how to deal with high-energy scattering
  beyond the mean field approximation. We review some of the main steps of
  this theoretical progress, like the role of Lorentz invariance and unitarity
  requirements, the importance of discreteness and fluctuations of gluon
  numbers (Pomeron loops), the high-energy QCD/statistical physics
  correspondence and the consequences for the saturation scale, the scattering
  amplitude and other, also measurable, quantities.

\end{abstract}

\section{Introduction}

%In this work we present the recent theoretical advances with respect to high
%energy QCD. We focus on the amplitude for a quark-antiquark dipole scattering
%off a target (hadron/nucleus) at high collison energy. The dipole-target 
%understanding of the energy
%dependence of the dipole-hadron/nucelus scattering amplitude
%
% is
%
%
%scattering at high energy is the  

%We wish to understand how the amplitude of a quark-antiquark dipole scattering
%off a hadron/nucleus chnges with increasing collision energy. 

The high-energy scattering of a dipole off a nucleus/hadron in the {\em mean
  field approximation} is decribed by the BK-equation~\cite{BK1,*BK2}. The
main results following from the BK-equation are the so-called geometric
scaling behaviour of the scattering
amplitude~\cite{Iancu:2002tr,Mueller:2002zm,MP1,*MP2} and the, roughly,
powerlike energy dependence of the saturation
scale~\cite{Mueller:2002zm,MP1,*MP2} which are supported by the HERA
data~\cite{Stasto:2000er,Triantafyllopoulos:2002nz}.

Over the last few years, we have had real breakthroughs in our understanding
of high-energy scattering near the unitarity limit. Namely, we have understood
how to deal with small-$x$ dynamics at high energy {\em beyond the mean field
  approximation}, i.e., beyond the BK-equation. In this work, after briefly
introducing the known dynamics in the mean-field case, we discuss the main
steps of the recent theoretical progress as follows: We start with a
discussion of the first step beyond the mean field approximation, which was
done in Ref.~\cite{Mueller:2004se} by enforcing the BFKL evolution in the
scattering process to satisfy natural requirements as unitarity limits and
Lorentz invariance. The consequence was a correction to the saturation scale
and the breaking of the geometric scaling at high energies.  Then, we explain
the relation between high-energy QCD and statistical physics found in
Ref.~\cite{Iancu:2004es} which has clarified the physical picture of, and the
way to deal with, the dynamics beyond the BK-equation. We explain that {\em
  gluon number fluctuations} from one scattering event to another and the {\em
  discreteness} of gluon numbers, both ignored in the BK evolution and also in
the Balitsky-JIMWLK equations~\cite{CGC1,*CGC2}, lead to the breaking of the
geometric scaling and to the correction to the saturation scale, respectively.
In a next step we show the new evolution equations, the so-called Pomeron loop
equations~\cite{Mueller:2005ut,IT1,*IT2,Kovner:2005nq}, which include a new
element in the evolution, the Pomeron loop. Finally, we discuss the
possibility of phenomenological
implications~\cite{KSB1,*KSB2,Kozlov:2007wm,Hatta:2006hs,Iancu:2006uc,Dumitru:2007ew}
of the recent theoretical advances. (For further studies on the recent
theoretical advances (not discussed here) see also~\cite{DD1,*DD2,*DD3,*DD4,N1,*N2,B1,*B2,*B3,SB1,*SB2,Bondarenko:2006rh,Blaizot:2006wp,Iancu:2006jw,Dumitru:2007ew,Munier:2006um,Ko1,*Ko2,*Ko3,Braun:2006gy}.)

%The study of
%dense-dilute~\cite{DD1,*DD2,*DD3,*DD4}, numerical~\cite{N1,*N2}, probability
%distribution~\cite{B1,*B2,*B3} Pomeron
%loops~\cite{SB1,*SB2,Bondarenko:2006rh,Blaizot:2006wp,Iancu:2006jw,Dumitru:2007ew,Munier:2006um,Ko1,*Ko2,*Ko3,Braun:2006gy}
%and their possible phenomenological
%consequences~\cite{KSB1,*KSB2,Kozlov:2007wm,Hatta:2006hs,Iancu:2006uc,Dumitru:2007ew}
%have atracteed a lot of interest over the last years.

\subsection{Mean field approximation}
\label{sec:mfa}

\begin{figure}
\begin{center}
\epsfig{file=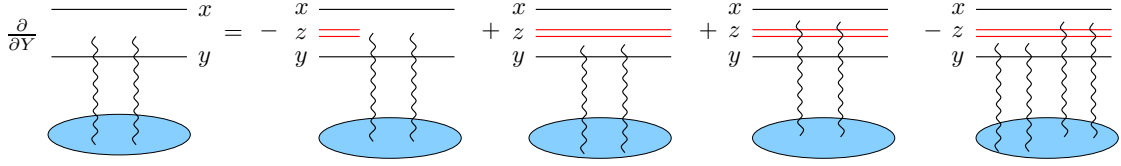,width=15cm}
\end{center}
\vspace*{-0.2cm}
\caption{The diagrammatic representation of the BK-equation for dipole-hadron
  scattering.}
\label{fig:BK_eq}
\end{figure}

Consider the high-energy scattering of a dipole of transverse size ${\bf
 r}=({\bf x}-{\bf y})$ off a target (hadron, nucleus) at rapidity $Y =
\ln(1/x)$. The $Y$-dependence of the $T$-matrix in the mean field
approximation is given by the BK-equation ($Y$-dependence is suppressed for
simplicity)
\begin{equation}
\frac{\partial T_{\bf xy}}{\partial Y} = \frac{\alpha_sN_c}{2\pi^2} \int
d^2{\bf z} \ M_{\bf
  xyz}\,\left[ -T_{\bf xy} + T_{\bf xz} + T_{\bf zy}-  T_{\bf xz}\,T_{\bf zy} \right] \ .
\label{eq:BK_eq}
\end{equation}
This equation can be interpreted as follows; If increasing the rapidity of the
dipole by $dY$ while keeping the rapidity of the target fixed, the probability
for the dipole to emit a gluon increases. In the large-$N_c$ limit the initial
quark-antiquark state plus the emitted gluon can be viewed as two dipoles -
one of the dipoles consists of the inital quark and the aniquark part of the
gluon whike the other dipoles is given by the quark part of the gluon and the
inital antiquark. The probability for the spliting of the inital dipole $({\bf
  x}-{\bf y})$ into two doughter dipoles with transverse sizes $({\bf x}-{\bf
  z})$ and $({\bf z}-{\bf y})$ is given by the weight in Eq.(\ref{eq:BK_eq}),
$\alpha_s N_c M_{\bf xyz}/(2\pi^2) d^2{\bf z} dY$, where ${\bf z}$ is the
transverse size of the emitted gluon and $M_{\bf xyz} = ({\bf x}-{\bf
  y})^2/[({\bf x}-{\bf z})^2 \, ({\bf x}-{\bf z})^2]$~\cite{Mueller:1993rr}. On the right-hand side
of Eq.(\ref{eq:BK_eq}), the first three terms (first one is virtual) describe
the scattering of a single dipole with the target whereas the last term gives
the simultanous scattering of the two doughter dipoles with the target, as
shown in Fig.~\ref{fig:BK_eq}. Without the last term, the BK-equation reduces
to a linear equation, the BFKL equation, which gives the growth of $T_{\bf
  xy}$ with rapidity, while the nonlinaer term, $T_{xz}\,T_{zy}$, tames the
growth of $T_{\bf xy}$ such that the unitarity limit, $T_{\bf xy} \leq 1$, is
satisfied.

One of the main results following from the BK-equation is the {\em geometric
  scaling} behaviour of the
$T$-matrix~\cite{Iancu:2002tr,Mueller:2002zm,MP1,*MP2} in a large kinematical
window
\begin{equation}
T(r,Y) = T(r^2\,Q^2_s(Y)) \ , 
\label{eq:gs}
\end{equation}
where $Q_s(Y)$ is the so-called {\em saturation momentum} defined such that
$T(r \simeq 1/Q_s,Y)$ be of ${\cal{O}}(1)$.  Eq.~(\ref{eq:gs}) means that the
$T$-matrix scales with a single quantity $r^2\,Q^2_s(Y)$ rather than depending
on $r$ and $Y$ separatelly. This behaviour implies a similar scaling for the
DIS cross section, $\sigma^{\gamma^{*}p}(Y,Q^2) =
\sigma^{\gamma^{*}p}(Q^2/Q^2_s(Y))$, which is supported by the HERA
data~\cite{Stasto:2000er}.

Another important result that can be extracted from the BK-equation is the
rapidity dependence of the saturation momentum (leading-$Y$
contribution)\cite{Mueller:2002zm,MP1,*MP2},
\begin{equation} 
Q^2_s(Y) = Q_0^2 \ \mbox{Exp}\left[\frac{2\alpha_s N_c}{\pi}
\frac{\chi(\lambda_0)}{1-\lambda_0} Y \right] \ ,  
\label{eq:Qrmf}
\end{equation}
where $\chi(\lambda)$ is the BFKL kernel and $\lambda_0 = 0.372$.  

%The
%exponent in (\ref{eq:Qrmf}) is known at NLO accuraccy and has the value
%$\lambda \simeq 0.3$ which is in good agreement with values obtained from fits
%to the HERA data.

The shape of the $T$-matrix resulting from the BK-equation is preserved in the
transition region from weak ($T\simeq \alpha_s^2$) to strong ($T \simeq 1$)
scattering with increasing $Y$, showing a ``travelling wave'' behaviour as
sketched in Fig.\ref{fig:BK_eq1}, on the left hand side. With increasing $Y$,
the saturation region at $r \gg 1/Q_s(Y)$ where $T \simeq 1$ however widens
up, including smaller and smaller dipoles, due to the growth of the saturation
momentum. As we will see in the next sections, the situation changes a lot
once gluon number fluctuations are taken into account.

\section{Beyond the mean field approximation}
\label{sec:bmf}

\subsubsection{Lorentz invariance and unitarity requirements}

Let's start with an elementary dipole of size $r_1$ at rapidity $y = 0$ and
evolve it using the BFKL evolution up to $y=Y$. The number density of dipoles
of size $r_2$ at $Y$ in this dipole, $n(r_1,r_2,Y)$, obeys a completeness
relation
\begin{equation}
n(r_1,r_2,Y) = \int \frac{d^2 r}{2\pi r^2} \ n(r_1,r,Y/2) \ n(r,r_2,Y/2) \ 
\label{eq:nlf}
\end{equation}
where on the right hand side the rapidity evolution is separated in two
successive steps, $y=0 \to y=Y/2 \to y=Y$. With
\begin{equation}
T(r_1,r_2,Y) \simeq c \ \alpha_s^2 \ r_2^2 \ n(r_1,r_2,Y) 
\label{eq:Tn}
\end{equation}
eq.(\ref{eq:nlf}) can be approximately rewritten in terms of the $T$-matrix as
\begin{equation} 
  \left(\frac{1}{r_2^2} \ T(r_1,r_2,Y)\right) \simeq \ \frac{1}{2 c \alpha_s^2}\ \int d\rho \
  \ \left ( \frac{1}{r^2} \ T(r_1,r,Y/2) \right) \ \left( \frac{1}{r_2^2} \ 
    T(r,r_2,Y/2) \right) 
\label{eq:Tlf}
\end{equation}
where $\rho = \ln(r^2_0/r^2)$. In Ref.~\cite{Mueller:2004se} it was realized
that the above completeness relations, or, equivalently, the Lorentz
invariance, is satisfied by the BK evolution only by violating unitarity
limits. This can be illustrated as follows: Suppose that $r_2$ is close to the
saturation line, $r_2 \simeq 1/Q_s(Y)$, so that the left hand side of
Eq.(\ref{eq:Tlf}) is large.  On the right hand side of Eq.(\ref{eq:Tlf}) it
turns out that $T(r_1,r,Y/2)/r^2$ is typically very small in the region of
$\rho$ which dominates the integral. This means that $T(r,r_2,Y/2)/r_2^2$ must
be typically very large and must violate unitarity, $T(r,r_2,Y/2) \gg 1$, in
order to satisfy (\ref{eq:Tlf}).

The simple procedure used in Ref.~\cite{Mueller:2004se} to solve the above
problem was to limit the region of the $\rho$-integration in Eq.(\ref{eq:Tlf})
by a boundary $\rho_2(Y/2)$ so that $T(r,r_2,Y/2)/r_2^2$ would never violate
unitarity, or $T(r_1,r,Y/2)/r^2$ would always be larger than $\alpha_s^2$. 
The main consequence of this procedure, i.e., BK evolution plus boundary
correcting it in the weak scattering region, is the following scaling behaviour of
the $T$-matrix near the unitarity limit
\begin{equation}
T(r,Y) = T\left(\frac{\ln(r^2Q_s^2(Y))}{\alpha_s Y/(\Delta\rho)^3}\right)
\label{eq:Trbmf}
\end{equation}
and the following energy dependence of the saturation momentum
\begin{equation}
Q_s^2(Y) = Q_0^2 \ \mbox{Exp}\!\left[\frac{2\alpha_s N_c}{\pi}
\frac{\chi(\lambda_0)}{1-\lambda_0} Y \left(1-\frac{\pi^2
    \chi''(\lambda_0)}{2(\Delta\rho)^2 \chi(\lambda_0)}\right)\right]
\label{eq:Qrbmf}
\end{equation}
with
\begin{equation}
\Delta\rho = \frac{1}{1-\lambda_0} \ln\frac{1}{\alpha_s^2} +
\frac{3}{1-\lambda_0} \ln \ln \frac{1}{\alpha_s^2} + \mbox{const.} \ .
\end{equation}
Eq.(\ref{eq:Trbmf}) shows the breaking of the geometric scaling, which was the
hallmark of the BK-equation shown in Eq.(\ref{eq:gs}), and Eq.(\ref{eq:Qrbmf})
shows the correction to the saturation momentum (cf. Eq.(\ref{eq:Qrmf})), both
emerging as a consequence of the evolution beyond the mean field approximation.

%This simple procedure used in Ref.~\cite{Mueller:2004se} is the first step in
%trying to incorporate Lorentz boost invariance in small-$x$ evolution, which
%is missed by the BK-equation or the Balitsky-JIMWLK hierarchy~\cite{IV1,*IV2}
%in intermediate steps of the evolution. 
%It turnes out that the result obtained for
%the saturation momentum, Eq.(\ref{eq:Qrbmf}), is general while the result for
%the geometric scaling violations, Eq.(\ref{eq:Trbmf}), misses some aspects of
%the stochastic evolution and will be corrected somewhat, as explained in the
%next section. 
%
%\begin{figure}
%\begin{center}
%\epsfig{file=abs_bound_ite.eps,width=8cm}
%\end{center}
%\caption{..}
%\end{figure}
%
\begin{figure}
\begin{center}
\epsfig{file=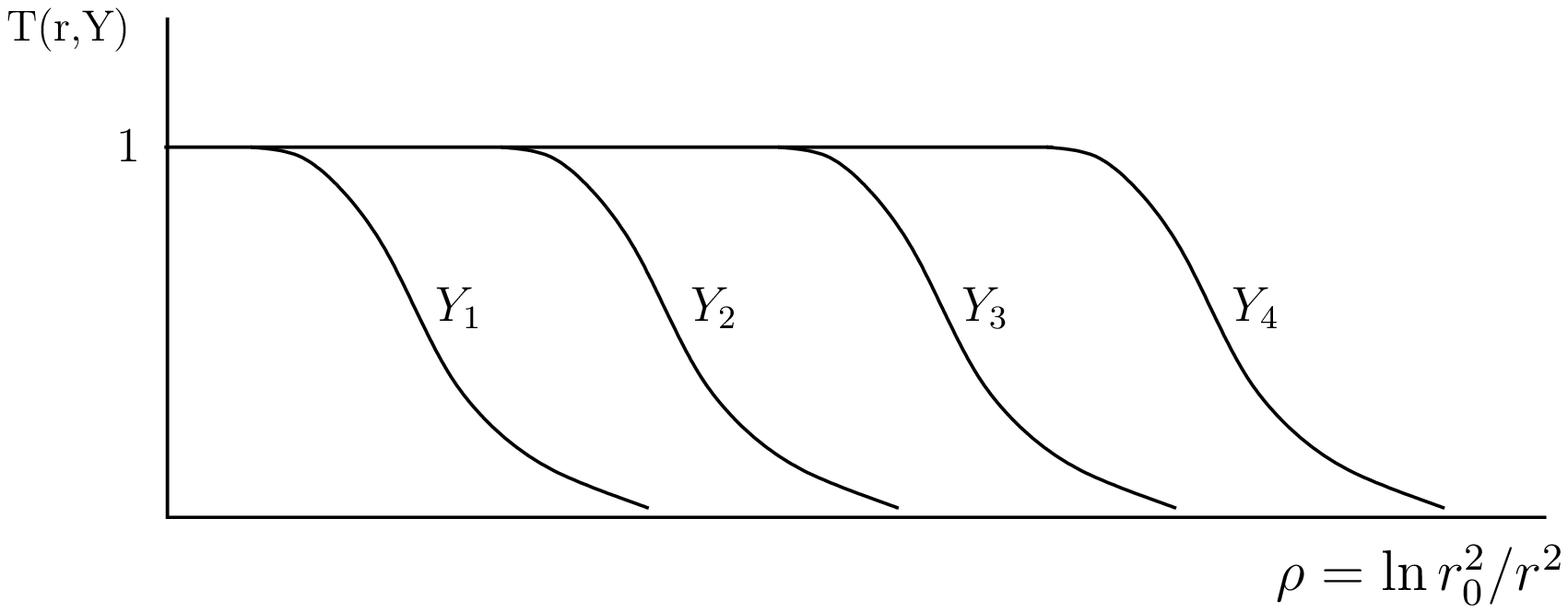,width=7cm} \hfill 
\epsfig{file=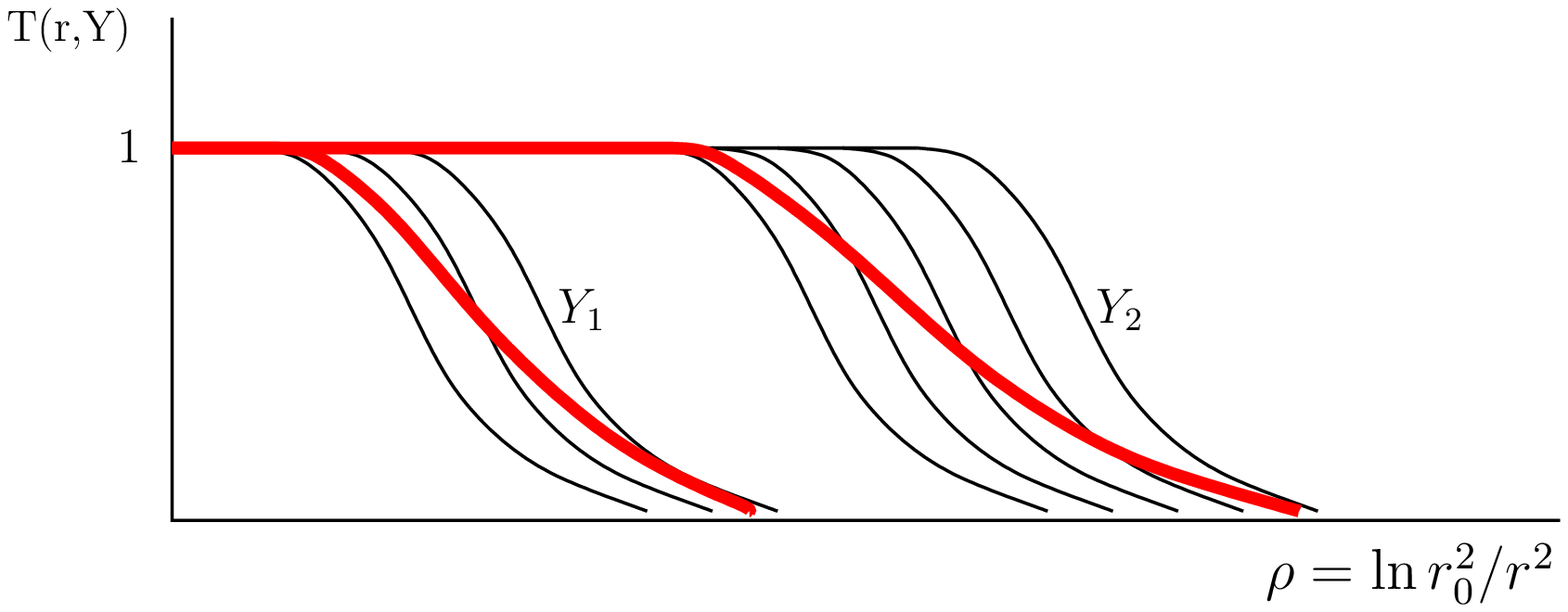,width=7cm} 
\end{center}
\caption{Left-hand side: The ``travelling wave'' behaviour of the solution to
  the BK-equation. Right-hand side: The $T$-matrix at two different
  rapidities, $Y_1$ and $Y_2$, for different events (thin lines). The thick
  lines represent the average over the events, $\langle T \rangle$, at the two
  rapidities, respectivelly. The shape of $\langle T \rangle$ becomes flatter 
  with rising rapidity.}
\label{fig:BK_eq1}
\end{figure}
\subsubsection{Statistical physics - high density QCD correspondence}
\label{sec:smqcd}
%
%Another access to small-$x$ physics beyond teh mean field approximation,
%inspired by the dynamics of reaction-diffusion processes in statistical
%physics, was outlined in Ref.~~\cite{Iancu:2004es}.
The high energy evolution can be viewed also in another way which is inspired
by dynamics of reaction-diffusion processes in statistical
physics~\cite{Iancu:2004es}. To show it, let's consider an elementary target
dipole of size $r_1$ which evolves from $y=0$ up to $y=Y$ and is then probed
by an elementary dipole of size $r$, giving the amplitude $\bar{T}(r_1,r,Y)$.
It has become clear that the evolution of the target dipole is {\em
  stochastic} leading to random dipole number realizations inside the target
dipole at $Y$, corresponding to different events in an experiment.  The
physical amplitude, $\bar{T}(r_1,r,Y)$, is then given by averaging over all
possible dipole number realizations/events, $\bar{T}(r_1,r,Y) = \langle
T(r_1,r,Y)\rangle$, where $T(r_1,r,Y)$ is the amplitude for dipole $r$
scattering off a particular realization of the evolved target dipole at $Y$.
An illustration is shown in Fig.\ref{fig:BK_eq1}, the right-hand side plot,
where the $T$-matrix for different events (thin lines) and the average over
all events (thick lines), $\langle T \rangle$, are shown at two different
rapidities, respectivelly.

The mean field description breaks down at low target dipole occupancy due to
the {\em discreteness and the fluctuations of dipole numbers}. Because of
discreteness the dipole occupancy can not be less than one for any dipole
size.  Taking this fact into account by using the BK equation with a cutoff
when $T$ becomes of order $\alpha^2_s$~\cite{Iancu:2004es}, or the occupancy
of order one (see Eq.(\ref{eq:Tn})), leads exactly to the same correction for
the saturation momentum as given in Eq.(\ref{eq:Qrmf}).  The latter cutoff is
essentially the same as, and gives a natural explanation of, the boundary
used in Ref.\cite{Mueller:2004se} and briefly explained in the previous section.

%The physical amplitude is, of course, given by averaging over all possible
%gluon number realizations/events,
%%
%\begin{equation}
%\bar{T}(r_1,r_2,Y) = \langle T(r_1,r_2,Y)\rangle \ , 
%\end{equation}
%%
%where $T(r_1,r_2,Y)$ denotes the amplitude for dipole $r_2$ scattering
%off a particular realization of the evolved target dipole at $Y$. 

%The mean field decription should be a good approximation so long as the dipole
%occupancy is large compared to one in the evolved target dipole. At low dipole
%occupancy, the {\em discreteness} of the dipole number becomes important,
%especially the fact that dipole occupancy can not be less than one for any
%dipole size. The BK equation with a cutoff when $T$ becomes of size
%$\alpha^2_s$, or the occupancy of order one (see Eq.(\ref{})), leads to
%exactly the same correction for the saturation momentum as given in
%Eq.(\ref{}).  The latter cutoff is essentially the same as the boundary used
%in Ref.\cite{} and briefly explained in the previous section.

The dipole number fluctuations in the low dipole occupancy region result in
fluctuations of the saturation momentum from event to event, with the strength
\begin{equation}
\sigma^2 =  \langle \rho_s^2 \rangle
- \langle \rho_s \rangle^2 \propto   \frac{\alpha_s
  Y}{(\Delta\rho)^3} \ 
\label{eq:disp}
\end{equation}
extracted from numerical simulations of statistical models. The averaging over
all events with random saturation momenta, in order to get the physical
amplitude, causes the breaking of the geometric scaling and replaces it by a
new scaling law, the so-called {\em diffusive scaling}, in which case the
scattering amplitude is a function of another variable,
\begin{equation}
\langle T(r,Y) \rangle = f\left(\frac{\ln(r^2 Q_s^2(Y))}{\sqrt{\alpha_s
      Y/(\Delta\rho)^3}}\right) \ .
\label{eq:Tqs}
\end{equation}
This equation differs from Eq.(\ref{eq:Trbmf}) since Eq.(\ref{eq:Trbmf})
misses dipole number fluctuations. Note that because of the geometric scaling
violation, the result in Eq.(\ref{eq:Tqs}) changes the shape as the rapidity
increases, as illustrated in Fig.~\ref{fig:BK_eq1} (right-hand side) by the
decreasing slope of the thick line with growing rapidity, in contrast to the
solution to the BK-equation in Eq.(\ref{eq:gs}).

\begin{figure}[htp]
\begin{center}
\epsfig{file=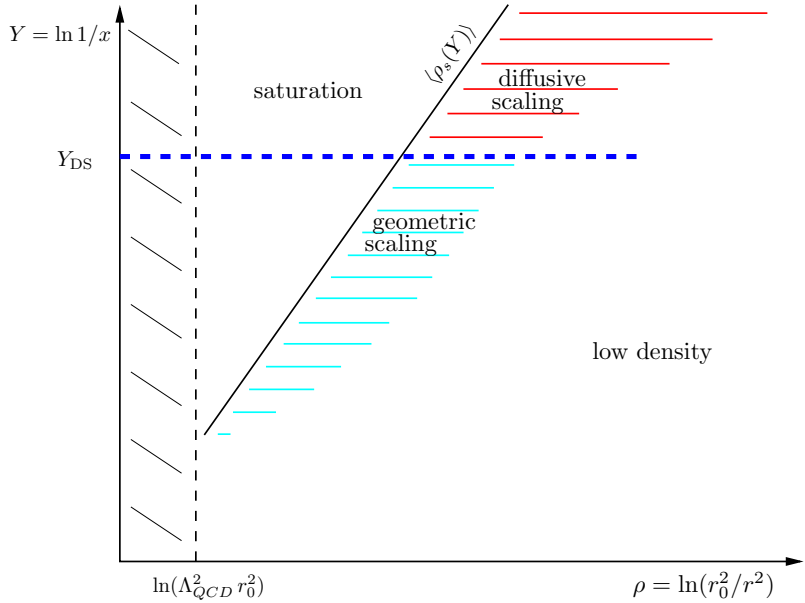,width=11cm} 
\end{center}
\caption{The phase diagram of the wavefunction of a highly evolved hadron.}
\label{fig:picture}
\end{figure}

The statistical physics/high-density QCD correspondence suggests the following
picture for the wavefunction of a highly evolved hadron which is probed by a
dipole of transverse size $r$: As the hadron is boosted to high rapidities the
density of gluons inside the hadron grows. Also the fluctuation in gluon
numbers, which is characterized by the dispersion in Eq.~(\ref{eq:disp}),
grows with rising rapidity. However, as long as $\sigma^2 \ll 1$, which means
$Y \ll Y_{DS} \simeq \ (\Delta \rho)^3/\alpha_s$, the effects of fluctuations
can be neglected and the evolution of the hadron is described to a good
approximation by the BK-equation. Thus, for $Y \ll Y_{DS}$, as shown in
Fig.\ref{fig:picture}, to the left of the saturation line, $\rho \ll \langle
\rho_s(Y)\rangle = \langle \ln(Q_s^2(Y)\,r_0^2)\rangle$, is the ``saturation
region'' with the ``large-size'' (small momentum) gluons at a large
density, of order $1/\alpha_s$ or the $T\simeq 1$, while the shadowed region
is the transition region from high to low gluon density, or the front of the at
$T$-matrix (geometric scaling regime). At higher rapidities, $Y \gg Y_{DS}$,
where the fluctuations become important, the geometric scaling regime is
replaced by the diffusive scaling given in Eq.~(\ref{eq:Tqs}).  

\subsubsection{Pomeron loop equations}
\begin{figure}
\begin{center}
\epsfig{file=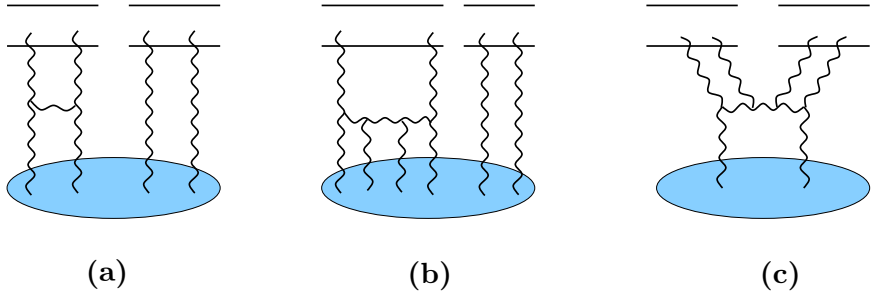,width=12cm}
\end{center}
\caption{Two dipoles scattering off a hadron: (a) BFKL interaction, (b) ``pomeron mergings'', (c) ``pomeron splittings''. }
\label{fig:BJIMWLK}
\end{figure}

It was always clear that the BK equation does not include fluctuations.
However, it took some time to realize that also the Balitsky-JIMWLK equations
do miss them. It turned out~(see first Reference in \cite{IT1,*IT2}) that the
Balitsky-JIMWLK equations do include BFKL evolution, ``pomeron mergings'' but
not also ``pomeron splittings'', which are represented by the three graphs in
Fig.~\ref{fig:BJIMWLK} for two dipoles scattering off a target,
respectivelly.  After this insight, the so-called Pomeron loop
equations~\cite{Mueller:2005ut,IT1,*IT2} have been constructed to account for
``pomeron splittings'' or dipole number fluctuations.

The Pomeron loop equations can be expressed in a Hamiltonian language, in
which case one extends the JIMWLK-equation~\cite{Mueller:2005ut}, or in terms
of scattering amplitudes~\cite{IT1,*IT2}, in which case the Balitsky equations
are extended. In order to be close to the BK-equation discussed in
sec.~\ref{sec:mfa}, we show the Pomeron loops using the scattering amplitude.
In the large-$N_c$ limit, they can be written either as a stochastic equation
of Langevin-type,
\begin{eqnarray} 
\frac{\partial T_{\bf xy}}{\partial Y} &=& \frac{\alpha_sN_c}{2\pi^2}
\int\limits_{\bf z} \ M_{\bf
  xyz}\,\left[ -T_{\bf xy} + T_{\bf xz} + T_{\bf zy}-  T_{\bf xz}\,T_{\bf zy}
\right] \nonumber \\ 
&+& \frac{\alpha_s}{2 \pi} \sqrt{\frac{\alpha_s N_c}{2\pi^2}}
\int\limits_{{\bf u,v,z}} {\cal A}({\bf x},{\bf y}|{\bf u},{\bf z}) \frac{|{\bf u}-{\bf v}|}{({\bf
    u}-{\bf z})^2} \sqrt{\nabla^2_{\bf u}\nabla^2_{\bf v}\,T_{\bf uv}}\ \nu({\bf u},{\bf v},{\bf z};Y)
\label{eq:Le}
\end{eqnarray}
%
%\begin{equation}
%\partial_Y T  = \alpha_s \left[T  - T T + \alpha_s\ \sqrt{T} \ \nu\right] 
%\label{eq:Le}
%\end{equation}
%
or, equivalently, as a hierarchy of coupled equations of averaged amplitudes,
where for simplicity we show only the first two of them, which read
\begin{eqnarray}
\frac{\partial \langle T_{\bf xy}\rangle}{\partial Y} &=& \frac{\alpha_sN_c}{2\pi^2}
\int\limits_{\bf z} \ M_{\bf
  xyz}\,\left[ -\langle T_{\bf xy}\rangle + \langle T_{\bf xz}\rangle +
  \langle T_{\bf zy}\rangle -  \langle T_{\bf xz}\,T_{\bf zy}\rangle
\right] \nonumber \\
\frac{\partial \langle T_{\bf xz} T_{\bf zy}\rangle}{\partial Y}&=& \frac{\alpha_sN_c}{2\pi^2}
\int\limits_{{\bf t}} \ M_{{\bf
  xzt}}\,\left[ -\langle T_{\bf xz}T_{\bf zy}\rangle + \langle T_{\bf xt}T_{\bf zy}\rangle +
  \langle T_{\bf tz}T_{\bf zy}\rangle -  \langle T_{\bf xt}\,T_{\bf tz}T_{\bf zy}\rangle
\right] \nonumber \\
&+& \frac{\alpha_sN_c}{2\pi^2}
\int\limits_{\bf t} \ M_{\bf
  zyt}\,\left[ -\langle T_{\bf xz}T_{\bf zy}\rangle + \langle T_{\bf xz}T_{\bf zt}\rangle +
  \langle T_{\bf xz}T_{\bf ty}\rangle -  \langle T_{\bf xz}\,T_{\bf zt}T_{\bf ty}\rangle
\right] \nonumber \\
&+& \left(\frac{\alpha_s}{2 \pi}\right)^2 \frac{\alpha_s N_c}{2\pi^2}\
\int\limits_{{\bf u},{\bf v}} {\cal R}({\bf x},{\bf z},{\bf z},{\bf y}|{\bf
  u},{\bf v}) \ \langle T_{\bf uv} \rangle
\label{eq:hie}
\end{eqnarray}
%
%
%\begin{eqnarray}
%\partial_Y \langle T \rangle &=& 
%\alpha_s \left[ \langle T \rangle  - \langle T T \rangle \right]\nonumber \\
%\partial_Y \langle T T \rangle &=&  
%\alpha_s \left[ \langle T T \rangle - \langle T\,T\,T \rangle + 
%\alpha^2_s \langle T \rangle \right] \ .
%\label{eq:hie}
%\end{eqnarray}
%
where the noise is non-diagonal (non-Gaussian) in the first two arguments
\begin{equation}
\langle \nu({\bf u}_1,{\bf v}_1,{\bf z}_1;Y)\nu({\bf u}_2,{\bf v}_2,{\bf
  z}_2;Y^{\prime}) \rangle = \delta_{{\bf u}_1{\bf v}_2} \ \delta_{{\bf
    u}_2{\bf v}_1} \ \delta_{{\bf z}_1{\bf z}_2} \ \delta_{YY^{\prime}} \ 
\end{equation}
the triple Pomeron kernel~\cite{Braun:1997nu} reads
\begin{equation}
{\cal R}({\bf x}_1,{\bf y}_1,{\bf x}_2,{\bf y}_2|{\bf u},{\bf v}) =
\int\limits_{\bf z} \nabla^2_{\bf u}\nabla^2_{\bf v}\, \left[M_{\bf uvz}\,{\cal A}({\bf x}_1,{\bf y}_1|{\bf u},{\bf z})\,{\cal A}({\bf x}_2,{\bf y}_2|{\bf z},{\bf v})\right] 
\end{equation}
and $\alpha_s^2 {\cal A}$ is the amplitude for dipole-dipole scattering in the
two-gluon exchange approximation and for large-$N_c$, with 
\begin{equation}
{\cal A}({\bf x},{\bf y}|{\bf u},{\bf v}) = \frac{1}{8} \ \ln^2
\left[\frac{({\bf x}-{\bf v})^2\,({\bf y}-{\bf u})^2}{({\bf x}-{\bf
      u})^2\,({\bf y}-{\bf v})^2}\right] \ .
\end{equation} 
In above equations the integrations are always over transverse sizes,
$\int_{{\bf x},{y}} = d^2 x \ d^2y$.  

Last term in Eq.(\ref{eq:Le}), containing the non-Gaussian noise $\nu$, is
new as compared with the BK-equation and accounts for fluctuations in the
dipole numbers or the stochastic nature of the evolution in small-$x$ physics.
The hierarchy in Eq.(\ref{eq:hie}) reduces to the BK-equation only in the mean
field approximation, i.e., when $\langle T\,T \rangle = \langle T\rangle
\langle T\rangle$. The hierarchy in Eq.(\ref{eq:hie}), as compared with the
Balitsky-JIMWLK hierarchy, involves in addition to linear BFKL evolution
(Fig.\ref{fig:BJIMWLK}(a)) and pomeron mergings (Fig.\ref{fig:BJIMWLK}(b)),
also pomeron splittings (Fig.\ref{fig:BJIMWLK}(c)), and therefore, in the
course of the evolution, also {\em pomeron loops}. The three pieces of
evolution are represented by the linear terms, nonlinear terms and the last
term on the right-hand side of the second equation in Eq.(\ref{eq:hie}),
respectivelly, which describes the scattering of two dipoles off a target.

%A diagrammatic
%representation of the linear BFKl evolution, pomeron mergings and pomeron
%splitting is shown in Fig.\ref{BJIMWLK}(a), Fig.\ref{BJIMWLK}(b) and
%Fig.\ref{BJIMWLK}(c), respectivelly. 

%
%

%The complexity of the present evolution equations has hindered  analytic, as
%well as numeric, solutionThe present evolution equations have not been solved so farDue to the complexity of the evolution equations%

\subsubsection{Phenomenology}

It isn't yet clear at which energy fluctuation/Pomeron loop effects start
becoming important. The results shown in the previous sections,
Eq.(\ref{eq:Qrbmf}) and Eq.(\ref{eq:Tqs}), are valid at asymptotic energies. A
solution to the evolution equations, which is not yet available because of
their complexity, would have helped to better understand the subasymptotics.

Using the statistical physics/high density QCD correspondence,
phenomenological consequences of fluctuations in the fixed coupling case have
been studied, for example for DIS and diffractive cross
sections~\cite{Hatta:2006hs}, forward gluon production in hadron-hadron
collisions~\cite{Iancu:2006uc} and for the nuclear modification factor
$R_pA$~\cite{KSB1,*KSB2}, in case fluctuations become important in the range
of LHC energies. Recently, in the fixed coupling case, it has been shown that
dipole-proton scattering amplitudes which include fluctuation effects seem to
describe better the HERA data.  Also the parameters turn out reasonable: The
diffusion coefficient $D \simeq 0.35$ ($\sigma^2 = D\,Y$) is in agreement with
numerical simulations of approximations to Pomeron loop
equations~\cite{N1,Iancu:2006jw} and the saturation exponent $\lambda \simeq
0.2$ ($Q_s^2 = (x_0/x)^{\lambda}$) is decreased as expected theoretically. On
the other hand, allowing the coupling to run, however, within a toy
model~\cite{Iancu:2006jw} which is supposed to mimic the QCD evolution
equations with Pomeron loops, it has been argued that gluon number
fluctuations/pomeron loops can be neglected in the range of HERA and LHC
energies. See also Refs.~\cite{Mueller:2004se,Beuf:2007qa} for more studies on
running coupling plus gluon number fluctuations.

%The new scaling behaviour
%(diffusive scaling) for DIS cross sections, diffractive cross sections and
%multiplicities in hadron-hadron collisions has been studied in~\cite{} and a
%new phenomenon originating from fluctuations, ``total gluon shadowing''for
%$R_pA$ in the fixed coupling, was found in~\cite{}. 

%------------------------------------------------------------------------------
%       Bibliography
%------------------------------------------------------------------------------
\begin{footnotesize}
\bibliographystyle{blois07} 
{\raggedright
\bibliography{blois07}
}
\end{footnotesize}
\end{document}